\documentclass{aastex}

\usepackage{emulateapj5}

\slugcomment{ApJ, in press}
\shorttitle{Radio survey of GRB hosts}
\shortauthors{Micha{\l}owski et al.}
\newcommand{\myemail}{mm@roe.ac.uk}
\newcommand{\tablefootmark}[1]{\tablenotemark{{\it #1}}}
\newcommand{\tablefoottext}[2]{\tablenotetext{#1}{#2}}
\newcommand{\tablefoot}{\tablecomments}
\newcommand{\urltt}[1]{\url{\texttt{#1}}}

\usepackage{color}

\newcommand{\GiveRef}[1]{\citetalias{#1}: \citet{#1}}
\newcommand{\msun}{\mbox{$M_\odot$}}
\newcommand{\msunyr}{\mbox{\msun\,yr$^{-1}$}}
\newcommand{\inst}[1]{\altaffilmark{#1}}
\usepackage{comment}

\begin{document}

\title{The optically unbiased GRB host (TOUGH) survey. VI. Radio Observations  at $z\lesssim1$ and consistency with typical star-forming galaxies$^1$%
}

\author{M.~J.~Micha{\l}owski\inst{\ref{inst:mich},\ref{inst:dark}}, 
	A.~Kamble\inst{\ref{inst:atish}},
	J.~Hjorth\inst{\ref{inst:dark}}, 
	D.~Malesani\inst{\ref{inst:dark}},
	R.~F.~Reinfrank\inst{\ref{inst:rob1},\ref{inst:rob2}},
	L.~Bonavera\inst{\ref{inst:bona}},
	J.~M.~Castro~Cer\'{o}n\inst{\ref{inst:jm}},
	E.~Ibar\inst{\ref{inst:ibar}},
	J.~S.~Dunlop\inst{\ref{inst:mich}},
	J.~P.~U.~Fynbo\inst{\ref{inst:dark}},
	M.~A.~Garrett\inst{\ref{inst:gar1},\ref{inst:gar2},\ref{inst:gar3}},
	P.~Jakobsson\inst{\ref{inst:palle}},
	D.~L.~Kaplan\inst{\ref{inst:atish}},
	T.~Kr\"{u}hler\inst{\ref{inst:dark}},
	A.~J.~Levan\inst{\ref{inst:lev}},
	M.~Massardi\inst{\ref{inst:mas}},
	S.~Pal\inst{\ref{inst:pal}},
	J.~Sollerman\inst{\ref{inst:soll}},
	N.~R.~Tanvir\inst{\ref{inst:tan}},
	A.~J.~van~der~Horst\inst{\ref{inst:horst}},
	D.~Watson\inst{\ref{inst:dark}},
and	K.~Wiersema\inst{\ref{inst:tan}}
	}

\altaffiltext{1}{Based on observations collected at the European Southern Observatory, Paranal, Chile (ESO Large Programme 177.A-0591), the Australian Telescope Compact Array, the Giant Metrewave Radio Telescope, the Very Large Array and the Westerbork Synthesis Radio Telescope.}
\altaffiltext{2}%
{SUPA 
{(Scottish Universities Physics Alliance)}, Institute for Astronomy, University of Edinburgh, Royal Observatory, Edinburgh, EH9 3HJ, UK; \myemail  \label{inst:mich}}
\altaffiltext{3}%
{Dark Cosmology Centre, Niels Bohr Institute, University of Copenhagen, Juliane Maries Vej 30, DK-2100 Copenhagen \O, Denmark  \label{inst:dark}}
\altaffiltext{4}%
{Physics Department, University of  Wisconsin-Milwaukee, Milwaukee, WI 53211, USA \label{inst:atish}}
\altaffiltext{5}%
{CSIRO Astronomy and Space Science, P.O. Box 76, Epping, NSW 1710, Australia \label{inst:rob1}}
\altaffiltext{6}%
{School of Chemistry and Physics, The University of Adelaide, Adelaide, SA 5005, Australia\label{inst:rob2}}
\altaffiltext{7}%
{Instituto de F\'isica de Cantabria, CSIC-Universidad de Cantabria, Avda. de los Castros s/n, 39005 Santander, Spain\label{inst:bona}}
\altaffiltext{8}%
{Department of Radio Astronomy, Madrid Deep Space Communications Complex (INTA-NASA/INSA), Ctra.~M-531, km.~7, E-28.294 Robledo de Chavela (Madrid), Spain\label{inst:jm}}
\altaffiltext{9}%
{UK Astronomy Technology Centre, Royal Observatory, Blackford Hill, Edinburgh EH9 3HJ, UK  \label{inst:ibar}} 
\altaffiltext{10}%
{Netherlands Institute for Radio Astronomy (ASTRON), Postbus 2, 7990 AA Dwingeloo, The Netherlands\label{inst:gar1}}
\altaffiltext{11}%
{Leiden Observatory, University of Leiden, P.B. 9513, Leiden 2300 RA, The Netherlands\label{inst:gar2}}
\altaffiltext{12}%
{Centre for Astrophysics and Supercomputing, Swinburne University of Technology, Hawthorn, Victoria 3122, Australia \label{inst:gar3}}
\altaffiltext{13}%
{Centre for Astrophysics and Cosmology, Science Institute, University of Iceland, Dunhagi 5, 107 Reykjav\'{i}k, Iceland \label{inst:palle}}
\altaffiltext{14}%
{Department of Physics, University of Warwick, Coventry CV4 7AL, UK\label{inst:lev}}
\altaffiltext{15}%
{INAF-Istituto di Radioastronomia, via Gobetti 101, 40129, Bologna, Italy\label{inst:mas}}
\altaffiltext{16}%
{ICRAR, University of Western Australia, 35 Stirling Highway, Crawley, WA, Australia  \label{inst:pal} }
\altaffiltext{17}%
{The Oskar Klein Centre, Department of Astronomy, AlbaNova, Stockholm University, 106 91 Stockholm, Sweden \label{inst:soll}  }
\altaffiltext{18}%
{Department of Physics and Astronomy, University of Leicester, University Road, Leicester, LE1 7RH, UK \label{inst:tan} }
\altaffiltext{19}%
{Astronomical Institute ``Anton Pannekoek", University of Amsterdam, Science Park 904, 1098XH Amsterdam, The Netherlands \label{inst:horst} }
\email{\myemail}


\begin{abstract}

The objective of this paper is to determine the level of obscured star formation activity and dust attenuation in a sample of gamma-ray burst (GRB) hosts; and to test the hypothesis that GRB hosts have properties consistent with those of the general star-forming galaxy populations.
 We present a radio continuum survey of all $z<1$ GRB hosts in The Optically Unbiased GRB Host (TOUGH)  sample
supplemented with radio data for all (mostly pre-{\it Swift}) GRB-SN hosts discovered before 2006 October.
We  present new radio data for 22 objects and have obtained a detection for three of them (GRB 980425, 021211, 031203; none in the TOUGH sample), increasing the number of radio-detected GRB hosts from two to five. The star formation rate (SFR) for the GRB 021211 host of $\sim825\, M_\odot$ yr$^{-1}$, the highest ever reported for a GRB host, places it in the category of  ultraluminous infrared galaxies. We found that at least $\sim63$\% of GRB hosts have  SFR$\mbox{}<100\, M_\odot$ yr$^{-1}$ and at most $\sim8$\% can have SFR$\mbox{}>500\, M_\odot$ yr$^{-1}$. For the undetected hosts the mean radio flux 
($<35\,\mu$Jy $3\sigma$) corresponds to an average SFR$\mbox{}< 15\,M_\odot$ yr$^{-1}$.  Moreover, $\gtrsim88$\% of the $z\lesssim1$ GRB hosts have ultraviolet dust attenuation $A_{\rm UV}<6.7$ mag (visual attenuation $A_V<3$ mag). Hence we did not find evidence for large dust obscuration in a majority of GRB hosts.  Finally, we found that the distributions of SFRs and $A_{\rm UV}$ of GRB hosts are consistent with those of Lyman break galaxies, H$\alpha$ emitters at similar redshifts and of galaxies from cosmological simulations.
The similarity of the GRB population with other star-forming galaxies is consistent with the hypothesis that GRBs, a least at $z\lesssim1$,
trace a large fraction of all star formation, and are therefore less biased indicators than once thought.
\end{abstract}

\keywords{dust, extinction --- galaxies: evolution  --- galaxies: ISM --- galaxies: star formation  ---  gamma-ray burst: general --- radio continuum: galaxies}

\section{Introduction}

\setcounter{footnote}{18}

Long gamma-ray bursts (GRBs) mark the endpoint of the lives of very massive stars \citep[e.g.][]{hjorthnature, stanek} and due to the short life-times of such stars, they are believed to be excellent tracers of ongoing star formation in distant galaxies \citep{jakobsson05,yuksel08,kistler09,butler10, elliott12,robertson12}. However, before GRBs can be quantitatively used to trace the star formation history of the Universe, the properties of their host galaxies and the biases of the GRB samples must be understood. 

From optical/near-infrared studies we know that GRB hosts are often faint dwarf galaxies \citep{lefloch03,christensen04,savaglio09,castroceron10,levesque10,svensson10}.
 However, some host galaxies of (often optically-obscured) GRBs  are massive ($M_*\gtrsim10^{10.5}\,\msun$) and/or belong to the category of luminous infrared galaxies  (LIRGs; $L_{\rm IR}>10^{11}\,L_\odot$, or star formation rate SFR$\mbox{}\gtrsim17.2\,\msunyr$ using the conversion of \citealt{kennicutt}), e.g. GRB 980613 \citep{castroceron06,castroceron10}, 020127 \citep{berger07}, 020819B \citep{savaglio09,kupcuyoldas10}, 051022 \citep{castrotirado07,savaglio09}, 070306 \citep{jaunsen08,kruhler11},  080207 \citep{hunt11,svensson12}, 080325 \citep{hashimoto10}, and 080605 \citep{kruhler12}. 
Hence, the diversity of the GRB host sample is not yet fully described.

Moreover, short-wavelength emission does not give us a complete picture of GRB hosts, as it misses star formation that is heavily obscured by dust.
Unfortunately long-wavelength emission has been detected only in a handful of GRB hosts (\citealt{bergerkulkarni,berger,frail,tanvir,castroceron06,lefloch,lefloch12,priddey06,michalowski09,stanway10,hunt11,watson11,hatsukade12,svensson12,walter12}; see the compilation of submillimeter observations in \citealt{deugartepostigo12}), which hampers our ability to study GRB hosts in the context of galaxy evolution, as a significant fraction of star formation in the Universe is believed to be obscured by dust \citep[e.g.][]{lefloch05,chapman05,wall08,michalowski10smg}.

Here we attempt to improve this situation through an extensive multi-facility radio survey of GRB hosts limited to $z<1$ to obtain meaningful SFR limits,
drawn from The Optically Unbiased GRB Host \citep[TOUGH;][]{hjorth12} survey, which allows for unbiased statistical analysis.

Observations at radio wavelengths provide an unobscured (unaffected by dust) view on star-forming galaxies by tracking directly the recent \mbox{($\lesssim100$ Myr)} star formation activity through synchrotron radiation emitted by relativistic electrons accelerated by supernova (SN) remnants 
\citep{condon}.
Moreover, even though the radio emission accounts for only a fraction of the bolometric luminosity of a galaxy, it is well correlated with the infrared emission, a good tracer of both the  SFR and the dust mass in a galaxy.  
 Finally, the timescale probed by radio emission ($\lesssim100$ Myr) is much longer than the lifetime of a GRB progenitor \citep[$\sim5$--$8$ Myr;][]{sollerman05,hammer06,ostlin08,thone08}, 
so the radio emission probes the average star formation state of a galaxy, unlike a GRB rate, which measures the almost instantaneous star formation activity.

The objective of this paper is to (1) determine the level of obscured star formation activity and dust attenuation  in a representative sample of $z\lesssim1$ GRB hosts, and (2) test the hypothesis that GRB hosts are consistent with the general star-forming galaxy populations at similar redshifts.

We use a cosmological model with $H_0=70$ km s$^{-1}$ Mpc$^{-1}$,  $\Omega_\Lambda=0.7$, and $\Omega_{\rm m}=0.3$ and assume the \citet{salpeter} initial mass function  to which all estimates from the literature were converted to, if necessary.

\section{Sample}
\label{sec:sample}

\begin{table*}[!t!]
\scriptsize
\caption{Radio Observation Logs}
\label{tab:obslog}
\centering
\begin{tabular}{lllccccl}
\hline\hline
GRB & Array & Observation Dates & $t$\tablefootmark{a} & Frequency  & rms & Synth. Beam Size & Calibrators\tablefootmark{b}\\
& & & (hr) & (GHz) & ($\mu$Jy) & ($''$)\\

\hline
\multicolumn{8}{c}{GRB-SN subset}\\
\hline
\object{980425}\tablefootmark{c}\tablefootmark{d} & ATCA & 2007 Aug 18 & 9.00 & 4.8, 8.64 & 46, 27 & $76\times38$,  $37\times21$  & \object{PKS B1934-638}\\
\object{991208} & WSRT & 2007 Aug 2--3 & 11.97 & 1.43 & 47 & $14.5\times10.5$ & \object{3C286}, \object{3C48} \\
\object{020903} & GMRT & 2008 Jan 18--19, Mar 1 & 16.99 & 1.43 & 41 & $4.0\times2.0$ &  \object{3C48},  \object{2243-257}\\ 
\object{021211} & VLA & 2007 Jul 14 & 5.45 & 1.43 & 31 & $1.7\times1.5$ & \object{TXS 0542+498}, \object{PMN J0808+0514}\\ 
\object{031203}\tablefootmark{d}\tablefootmark{e} & ATCA & 2008 Jan 26 & 6.97 & 1.39, 2.37 & 46, 37 & $8.5\times3.4$, $6.3\times2.3$  &\object{PKS B1934-638}, \object{PKS B0826-373}\\
\object{041006} & GMRT & 2007 Aug 7--8 & 9.61 & 1.43 & 181 & $7.3\times2.0$ & \object{3C48},  \object{B2 0026+34}\\ 
\hline
\multicolumn{8}{c}{TOUGH $z<1$ unbiased subset}\\
\hline
\object{050416A} & WSRT & 2008 Apr 27--28 & 11.97 & 1.43 & 75 & $31.6 \times 9.6$ & \object{3C48}, \object{3C286}\\
\object{050525A} & WSRT & 2007 Aug 13--14 & 11.97 & 1.43 & 52 & $33.7 \times 14.8$ & \object{3C48}, \object{3C286}\\
\object{050824} & WSRT & 2007 Dec 26--27 & 11.97 & 1.43 & 100 & $39.9\times 15.5$ & \object{3C147}, \object{3C286}\\
\object{051016B} & WSRT & 2007 Dec 28--29 & 11.97 & 1.43 & 47 & $59.7 \times 14.1$ & \object{3C147}, \object{3C286} \\
\object{051117B}\tablefootmark{d}& ATCA & 2009 Aug 12 & 8.29 & 5.5, 9.0 & 12, 19 & $6.4\times1.7$, $3.9\times1.1$ & \object{PKS B1934-638}, \object{PKS B0607-157}\\
\object{060218} & WSRT & 2007 Aug 16--17 & 11.96 & 1.43 & 117 & $51.9 \times 15.2$ &  \object{3C48}, \object{3C286}\\
\object{060614}\tablefootmark{d}&  ATCA & 2009 Aug 9-10 & 8.84 & 5.5, 9.0 & 11, 14 &  $3.1\times1.9$, $1.7\times1.0$ & \object{PKS B1934-638} \\
\object{060729} & ATCA & 2008 Jan 26,28 & 11.36 & 1.39 & 35  & $7.4\times6.4$ & \object{PKS B1934-638}, \object{PKS B0515-674}\\
\object{060912A}\tablefootmark{f} & GMRT &  2009 Jun 1--2 & 10.00 & 1.43 &  $\cdots$ & $\cdots$ & \object{3C48} \\
\object{061021} & ATCA & 2008 Apr 18 & 7.90 & 1.39 & 36 & $20.0\times4.8$ & \object{PKS B1934-638}, \object{PKS B0919-260}\\
\object{061110A}\tablefootmark{f} & WSRT & 2007 Dec 29 & 12.00 & 1.43 & $\cdots$ & $\cdots$ & \object{3C147}, \object{3C286} \\
\object{070318} & ATCA & 2008 Apr 19 & 9.74 & 1.39 & 47 & $7.2\times4.2$ & \object{PKS B1934-638}, \object{PKS B0405-385}\\
\hline
\multicolumn{8}{c}{Other hosts}\\
\hline
\object{050915A} & ATCA & 2008 Jan 25,27 & 15.62 & 1.39 & 29 & $18.3\times5.5$ & \object{PKS B1934-638}, \object{PKS B0451-282}\\
\dots\tablefootmark{d}
& ATCA & 2011 Dec 19 & 9.78 & 5.5, 9.0 & 12, 15 & $5.9\times 2.1$, $3.7\times1.3$ & \object{PKS B1934-638}, \object{PKS B0537-286}\\
\object{060505}\tablefootmark{d}& ATCA & 2009  Aug 10--11 & 8.48 & 5.5, 9.0 & 17, 14  &  $5.2\times1.7$, $3.4\times1.1$ & \object{PKS B1934-638}, \object{PKS B2155-152 } \\
\dots 		          & GMRT & 2008 Jan 20--21 & 5.89 & 1.43 & 58  & $3.6\times2.3$ & \object{3C48},  \object{2243-257}\\ 
\object{060814} & WSRT & 2007 Dec 30  & 11.96 & 1.43 & 78 & $42.7 \times 14.9$ &  \object{3C48}, \object{3C286}\\
\object{070808} & GMRT & 2009 Jun 2--3 & 6.87 & 1.43 & 68 & $3.6\times2.2$ & \object{3C48},  \object{0022+002}\\ 
\hline
\end{tabular}
\tablefoot%
{
The horizontal lines divide the GRB-SN and the $z<1$ TOUGH subsets (see Section~\ref{sec:sample}) and the hosts which do not belong to any of these subsets. GRBs \object{050525A} and \object{060218} belong to the first two subsets.
} 
\tablefoottext{a}   {On-source integration time.}
\tablefoottext{b}   {The first (second) object was used as a primary (secondary) calibrator.  For WSRT the indicated objects were used as primary calibrators at the beginning and the end of the run.}
\tablefoottext{c}   {Data  published in \citet{michalowski09}.}
\tablefoottext{d}   {This object was observed simultaneously at two frequencies, see Table~\ref{tab:flux}.} 
\tablefoottext{e}   {Data  published in \citet{watson11}.}
\tablefoottext{f}   {Poor quality (interference and system malfunctions) of the data impedes the flux density measurement.}

\end{table*}

Our  target sample is composed of two subsets. The main  subset is drawn from the TOUGH sample based on the {\it Swift} satellite and a Very Large Telescope (VLT) Large Programme. 
 The survey design, the selection criteria, and the summary of the host properties (including redshifts) are presented in \citet{hjorth12}, the photometry and the host properties are analyzed in \citet{malesani12}, the redshifts are presented in \citet{jakobsson12} and \citet{kruhler12b}, and the Ly$\alpha$ properties are discussed in \citet{milvangjensen12}.  The reduced data  will be available from the TOUGH Web site\footnote{\urltt{http://www.dark-cosmology.dk/TOUGH}}. 
 The sample includes {\it all} long GRBs that exploded between 2005 March1 and 2007 August 10, observable from the southern hemisphere ($-70^\circ < \delta < +27^\circ$), with low Galactic foreground extinction ($A_V \le 0.5$~mag) and no bright star nearby, for which X-ray observations are available $<12$ hr after the burst (with $\le2''$ error circle radius) to allow the determination of  accurate positions.  Therefore, this X-ray-selected sample is constructed in a way that it is not biased against dusty systems and the selection does not depend on the host luminosity. We note that  the availability of redshift does depend on the host luminosity, but redshifts were measured for  $\sim77$\% (53/69) of {\it Swift}/VLT TOUGH GRBs \citep{hjorth12,jakobsson12}.
  Moreover, half of the TOUGH GRB redshifts were obtained from optical observations of afterglows, so the redshift recovery fraction is dependent on the host brightness only for the other half. Finally, fainter hosts (for which redshifts could not be measured) are less likely to be at $z<1$. Indeed $75$\% (12/16)  of TOUGH GRBs with unknown redshifts are fainter than $R=25$ mag, whereas the same is true only for $\sim17$\% (2/12) of GRBs confirmed to be at $z<1$. 
 We therefore conclude that the $z<1$ TOUGH sample has a completeness nearing $100$\%, and, in any case, larger than $\sim77$\%.

We restricted the TOUGH sample to $z<1$ to obtain meaningful radio constraints on SFRs.
The $z<1$ TOUGH unbiased  subset consists of 12 hosts 
of which all were observed within our program (see Table~\ref{tab:obslog} for observation logs). 
 \object{GRB 060814} at $z=1.92$,
\object{050915A} at $z=2.527$,
and \object{070808} with currently no redshift measurement
are  in our target sample, because they were initially believed to be at $z< 1$ \citep[e.g.][]{thone06gcn3}. 

The second subset includes (mostly pre-{\it Swift}) GRBs that were spectroscopically or photometrically confirmed  to be associated with SNe before 2006 October, namely the sample of \citet{ferrero06} plus \object{GRB 980425}/\object{SN 1998bw} \citep{galamanature} and \object{GRB 040924} \citep{soderberg06}. We targeted GRB-SN hosts, because their progenitors are securely established to be connected with recent star formation (see \citealt{hjorthsn} for a recent review of the GRB-SN connection). Since the detection of an SN component in a fading GRB afterglow is difficult at high redshifts, this selection imposes a practical limit of $z\lesssim1$. In total 15 hosts were selected (with an overlap of two hosts with the TOUGH subset)  of which eight were observed within our program and for the remaining seven the deep radio upper limits from the literature were adopted (see Table~\ref{tab:flux}).

In summary, our sample consists of 30 GRB hosts and we provide new radio observations for 22 of them. 

\section{Data}
\label{sec:data}

The radio data were collected using the Australian Telescope Compact Array (ATCA; proposals C1651, C1741, CX228) in the 6 km configuration (H168 for \object{GRB 980425}), the Giant Metrewave Radio Telescope  (GMRT; proposals 12MMc01, 13MMc01, 16\_093), the Very Large Array (VLA; proposal AM902) in A configuration, and the Westerbork Synthesis Radio Telescope  (WSRT; proposals R07B004, R08A002) in maxi-short configuration. The log of observations is presented in Table~\ref{tab:obslog}. 

\begin{figure*}[t]
\begin{center}
\includegraphics[width=\textwidth,, viewport=45 745 308 831,clip]{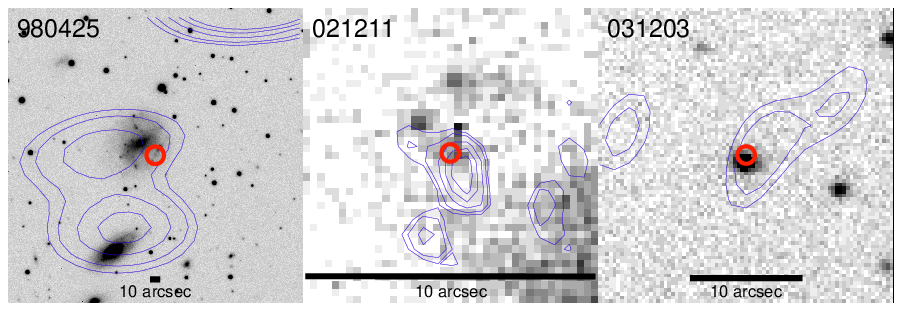}
\end{center}
\caption{Radio contours ({\it blue lines}) overlaid on the optical images of the detected GRB hosts. The size of each image depends on the host galaxy size and the resolution of the radio data: $3'$ for GRB 980425 ($4.8$ GHz), $10''$ for GRB 021211 ($1.43$ GHz), and $20''$ for GRB 031203 ($1.39$ GHz). 
The {\it red circles} (with arbitrary sizes) mark the position of GRBs (optical positions for GRB 980425 and 021211 from \citealt{fynbo00,fox03}, and X-ray position for GRB 031203 from \citealt{watson04b}).  The contours are $3$, $4$, $6$, $8$ and $10\sigma$ (see Table~\ref{tab:obslog}). North is up and east is left. 
 The optical data are from \citet[][GRB 980425]{sollerman05}, \citet[][GRB 021211]{dellavalle03}, and \citet[][GRB 031203]{mazzali06}.
}
\label{fig:image}
\end{figure*}

Data reduction and analysis were done using the MIRIAD  \citep{miriad} and AIPS\footnote{\urltt{http://www.aips.nrao.edu/cook.html}}  packages. Calibrated visibilities were Fourier transformed using ``robust'' or ``uniform'' weighting 
depending on which gave a better result for a particular  field. The resulting rms noise, beam sizes, and calibrators are listed in Table~\ref{tab:obslog} and the radio contours overlaid on the optical images for detected hosts are presented in Figure~\ref{fig:image}. 
 The data for the observations of GRB 060912A and 061110A were found to have been
severely affected by radio frequency interference and system malfunctions. Therefore,
we had to discard a significant fraction of the data and the remaining data were insufficient
to create reasonable radio images of the fields.

Flux densities were measured by fitting two-dimensional Gaussian functions to the region around the host and the errors were determined from the local rms on the images.
The hosts of GRBs \object{980425} and \object{031203} slightly overlap with radio objects $\sim70''$ south \citep[see][]{michalowski09} and $\sim6''$ northwest, respectively, so their flux densities were estimated by simultaneous fitting two Gaussian functions  with their centroids, sizes and orientations as free parameters. The lack of residuals left after the subtraction of these two Gaussians rules out a significant contamination of the nearby objects to the measured flux densities of the hosts.

\section{Results}
\label{sec:results}

\defcitealias{bloom01}{       1}
\defcitealias{tinney98}{       2}
\defcitealias{galama99gcn}{       3}
\defcitealias{hjorth00}{       4}
\defcitealias{castrotirado01}{       5}
\defcitealias{price02000911}{       6}
\defcitealias{price02}{       7}
\defcitealias{infante01gcn}{       8}
\defcitealias{price03}{       9}
\defcitealias{soderberg04b}{      10}
\defcitealias{vreeswijk06}{      11}
\defcitealias{hjorthnature}{      12}
\defcitealias{prochaska04}{      13}
\defcitealias{soderberg06}{      14}
\defcitealias{cenko05gcn}{      15}
\defcitealias{soderberg07}{      16}
\defcitealias{foley05gcn}{      17}
\defcitealias{fynbo05gcn}{      18}
\defcitealias{sollerman07}{      19}
\defcitealias{soderberg05gcn2}{      20}
\defcitealias{jakobsson12}{      21}
\defcitealias{pian06}{      22}
\defcitealias{price06gcn}{      23}
\defcitealias{thone06gcn}{      24}
\defcitealias{fynbo09}{      25}
\defcitealias{jaunsen07gcn}{      26}
\defcitealias{kruhler12b}{      27}
\defcitealias{ofek06gcn}{      28}
\defcitealias{salvaterra12}{      29}
\defcitealias{frail03}{      30}
\defcitealias{michalowski09}{      31}
\defcitealias{vreeswijk01radio}{      32}
\defcitealias{stanway10}{      33}
\defcitealias{berger}{      34}
\defcitealias{berger03}{      35}
\defcitealias{hatsukade12}{      36}
\defcitealias{fox03}{      37}
\defcitealias{vanderhorst05}{      38}
\defcitealias{watson11}{      39}
\defcitealias{wiersema08}{      40}
\defcitealias{soderberg04gcn}{      41}
\defcitealias{castroceron10}{      42}
\defcitealias{christensen04}{      43}
\defcitealias{savaglio09}{      44}
\defcitealias{svensson10}{      45}
\defcitealias{ovaldsen07}{      46}
\defcitealias{cano11}{      47}
\defcitealias{malesani12}{      48}
\begin{table*}[p]
\scriptsize
\caption{Radio Fluxes, Star formation Rates and Dust Attenuations of GRB Hosts}
\label{tab:flux}
\centering
\begin{tabular}{llc r@{ $\pm$ }l c r r@{ $\pm$ }l crr }
\hline\hline
GRB & \multicolumn{1}{c}{$z$} & Ref & \multicolumn{2}{c}{Flux Density} & Frequency & Ref & \multicolumn{2}{c}{SFR$_{\rm radio}$\tablefootmark{a}} & SFR$_{\rm UV}$\tablefootmark{b} & Ref\tablefootmark{c} & $A_{\rm V}$\tablefootmark{d} \\
& & & \multicolumn{2}{c}{($\mu$Jy)} & (GHz) & & \multicolumn{2}{c}{($M_\odot$ yr$^{-1}$)} & ($M_\odot$ yr$^{-1}$) & & (mag) \\
\hline
\multicolumn{12}{c}{ GRB-SN subset}\\
\hline
\object{970228} & 0.695 & \citetalias{bloom01} & \multicolumn{2}{c}{$<$ 69} & 
1.43 & \citetalias{frail03} & \multicolumn{2}{c}{$<$ 72} & 0.60 & 
\citetalias{castroceron10} & $<$2.3 \\ 
\object{980425} & 0.0085 & \citetalias{tinney98} & 420 & 50 & 4.80 & 
$\ddagger$,\citetalias{michalowski09} & 0.23 & 0.02 & 0.39 & 
\citetalias{castroceron10} & $\sim0$ \\ 
\dots & \dots & \dots & \multicolumn{2}{c}{$<$ 180} & 8.64 & 
$\ddagger$,\citetalias{michalowski09} & \multicolumn{2}{c}{$<$ 0.17} & 0.39 & 
\citetalias{castroceron10} & $\sim0$ \\ 
\object{990712} & 0.4337 & \citetalias{galama99gcn,hjorth00} & 
\multicolumn{2}{c}{$<$ 105} & 1.39 & \citetalias{vreeswijk01radio} & 
\multicolumn{2}{c}{$<$ 36} & 1.28 & \citetalias{christensen04} & $<$1.6 \\ 
\dots & \dots & \dots & \multicolumn{2}{c}{$<$ 36} & 5.50 & 
\citetalias{stanway10} & \multicolumn{2}{c}{$<$ 35} & 1.28 & 
\citetalias{christensen04} & $<$1.6 \\ 
\dots & \dots & \dots & \multicolumn{2}{c}{$<$ 129} & 9.00 & 
\citetalias{stanway10} & \multicolumn{2}{c}{$<$ 180} & 1.28 & 
\citetalias{christensen04} & $<$2.4 \\ 
\object{991208} & 0.7063 & \citetalias{castrotirado01} & 
\multicolumn{2}{c}{$<$ 32} & 1.43 & $\ddagger$ & \multicolumn{2}{c}{$<$ 35} & 
0.83 & \citetalias{christensen04} & $<$1.8 \\ 
\object{000911} & 1.058 & \citetalias{price02000911} & 
\multicolumn{2}{c}{$<$ 57} & 8.46 & \citetalias{berger} & 
\multicolumn{2}{c}{$<$ 608} & 1.40 & \citetalias{castroceron10} & $<$3.0 \\ 
\object{010921} & 0.451 & \citetalias{price02} & \multicolumn{2}{c}{$<$ 83} & 
1.43 & \citetalias{frail03} & \multicolumn{2}{c}{$<$ 32} & 1.60 & 
\citetalias{castroceron10} & $<$1.5 \\ 
\object{011121} & 0.36 & \citetalias{infante01gcn} & \multicolumn{2}{c}{$<$ 120}
 & 4.80 & \citetalias{frail03} & \multicolumn{2}{c}{$<$ 68} & 1.83 & 
\citetalias{savaglio09} & $<$1.8 \\ 
\object{020405} & 0.691 & \citetalias{price03} & \multicolumn{2}{c}{$<$ 42} & 
8.46 & \citetalias{berger03} & \multicolumn{2}{c}{$<$ 165} & 3.70 & 
\citetalias{castroceron10} & $<$1.9 \\ 
\object{020903} & 0.251 & \citetalias{soderberg04b} & \multicolumn{2}{c}{$<$ 53}
 & 1.43 & $\ddagger$ & \multicolumn{2}{c}{$<$ 5.39} & 0.42 & 
\citetalias{savaglio09} & $<$1.3 \\ 
\object{021211} & 1.006 & \citetalias{vreeswijk06} & 330 & 31 & 1.43 & 
$\ddagger$ & 825 & 77 & 0.72 & \citetalias{castroceron10} & 3.4 \\ 
\dots & \dots & \dots & \multicolumn{2}{c}{$<$ 34} & 2.10 & 
\citetalias{hatsukade12} & \multicolumn{2}{c}{$<$ 114} & 0.72 & 
\citetalias{castroceron10} & $<$2.5 \\ 
\dots & \dots & \dots & \multicolumn{2}{c}{$<$ 45} & 8.46 & \citetalias{fox03}
 & \multicolumn{2}{c}{$<$ 427} & 0.72 & \citetalias{castroceron10} & $<$3.1 \\ 
\object{030329} & 0.168 & \citetalias{hjorthnature} & 
\multicolumn{2}{c}{$<$ 420} & 1.40 & \citetalias{vanderhorst05} & 
\multicolumn{2}{c}{$<$ 17} & 0.14 & \citetalias{castroceron10} & $<$2.4 \\ 
\object{031203} & 0.105 & \citetalias{prochaska04} & 254 & 46 & 1.39 & 
$\ddagger$,\citetalias{watson11} & 3.83 & 0.69 & 4.30 & 
\citetalias{castroceron10} & $\sim0$ \\ 
\dots & \dots & \dots & 191 & 37 & 2.37 & $\ddagger$,\citetalias{watson11} & 
4.29 & 0.83 & 4.30 & \citetalias{castroceron10} & $\sim0$ \\ 
\dots & \dots & \dots & 216 & 50 & 5.50 & \citetalias{stanway10} & 9.13 & 2.11
 & 4.30 & \citetalias{castroceron10} & 0.4 \\ 
\dots & \dots & \dots & \multicolumn{2}{c}{$<$ 48} & 9.00 & 
\citetalias{stanway10} & \multicolumn{2}{c}{$<$ 3.09} & 4.30 & 
\citetalias{castroceron10} & $\sim0$ \\ 
\object{040924} & 0.859 & \citetalias{soderberg06} & \multicolumn{2}{c}{$<$ 63}
 & 4.90 & \citetalias{wiersema08} & \multicolumn{2}{c}{$<$ 274} & 
0.66\tablefootmark{e} & $\ddagger$,\citetalias{wiersema08} & $<$2.9 \\ 
\object{041006} & 0.716 & \citetalias{soderberg06} & \multicolumn{2}{c}{$<$ 45}
 & 2.10 & \citetalias{hatsukade12} & \multicolumn{2}{c}{$<$ 67} & 0.47 & 
\citetalias{savaglio09} & $<$2.4 \\ 
\dots & \dots & \dots & \multicolumn{2}{c}{$<$ 348} & 1.43 & $\ddagger$ & 
\multicolumn{2}{c}{$<$ 392} & 0.47 & \citetalias{savaglio09} & $<$3.3 \\ 
\dots & \dots & \dots & \multicolumn{2}{c}{$<$ 123} & 8.46 & 
\citetalias{soderberg04gcn} & \multicolumn{2}{c}{$<$ 525} & 0.47 & 
\citetalias{savaglio09} & $<$3.4 \\ 
\hline
\multicolumn{12}{c}{TOUGH $z<1$ unbiased subset}\\
\hline
\object{050416A} & 0.6528 & \citetalias{cenko05gcn,soderberg07} & 
\multicolumn{2}{c}{$<$ 447} & 1.43 & $\ddagger$ & \multicolumn{2}{c}{$<$ 405} & 
0.89 & \citetalias{savaglio09} & $<$3.0 \\ 
\object{050525A} & 0.606 & \citetalias{foley05gcn} & \multicolumn{2}{c}{$<$ 228}
 & 1.43 & $\ddagger$ & \multicolumn{2}{c}{$<$ 174} & 0.64 & 
\citetalias{castroceron10} & $<$2.7 \\ 
\object{050824} & 0.828 & \citetalias{fynbo05gcn,sollerman07} & 
\multicolumn{2}{c}{$<$ 111} & 1.43 & $\ddagger$ & \multicolumn{2}{c}{$<$ 175} & 
1.37 & \citetalias{svensson10} & $<$2.4 \\ 
\object{051016B} & 0.9364 & \citetalias{soderberg05gcn2} & 
\multicolumn{2}{c}{$<$ 220} & 1.43 & $\ddagger$ & \multicolumn{2}{c}{$<$ 465} & 
5.69 & $\ddagger$,\citetalias{ovaldsen07} & $<$2.2 \\ 
\object{051117B} & 0.481 & \citetalias{jakobsson12} & \multicolumn{2}{c}{$<$ 36}
 & 5.50 & $\ddagger$ & \multicolumn{2}{c}{$<$ 44} & 2.72 & 
$\ddagger$,\citetalias{ovaldsen07} & $<$1.4 \\ 
\dots & \dots & \dots & \multicolumn{2}{c}{$<$ 57} & 9.00 & $\ddagger$ & 
\multicolumn{2}{c}{$<$ 101} & 2.72 & $\ddagger$,\citetalias{ovaldsen07} & $<$1.8
 \\ 
\object{060218} & 0.0334 & \citetalias{pian06} & \multicolumn{2}{c}{$<$ 447} & 
1.43 & $\ddagger$ & \multicolumn{2}{c}{$<$ 1.00} & 0.05 & 
\citetalias{castroceron10} & $<$1.4 \\ 
\dots & \dots & \dots & \multicolumn{2}{c}{$<$ 117} & 5.50 & 
\citetalias{stanway10} & \multicolumn{2}{c}{$<$ 0.78} & 0.05 & 
\citetalias{castroceron10} & $<$1.3 \\ 
\dots & \dots & \dots & \multicolumn{2}{c}{$<$ 42} & 9.00 & 
\citetalias{stanway10} & \multicolumn{2}{c}{$<$ 0.48} & 0.05 & 
\citetalias{castroceron10} & $<$1.1 \\ 
\object{060614} & 0.125 & \citetalias{price06gcn} & \multicolumn{2}{c}{$<$ 33}
 & 5.50 & $\ddagger$ & \multicolumn{2}{c}{$<$ 2.35} & 0.02 & 
\citetalias{castroceron10} & $<$2.4 \\ 
\dots & \dots & \dots & \multicolumn{2}{c}{$<$ 42} & 9.00 & $\ddagger$ & 
\multicolumn{2}{c}{$<$ 3.72} & 0.02 & \citetalias{castroceron10} & $<$2.6 \\ 
\object{060729} & 0.54 & \citetalias{thone06gcn} & \multicolumn{2}{c}{$<$ 105}
 & 1.39 & $\ddagger$ & \multicolumn{2}{c}{$<$ 60} & 0.13 & \citetalias{cano11}
 & $<$3.0 \\ 
\object{061021} & 0.3463 & \citetalias{fynbo09} & \multicolumn{2}{c}{$<$ 108} & 
1.39 & $\ddagger$ & \multicolumn{2}{c}{$<$ 22} & 0.03\tablefootmark{e} & 
$\ddagger$,\citetalias{jakobsson12} & $<$3.2 \\ 
\object{070318} & 0.836 & \citetalias{jaunsen07gcn} & 
\multicolumn{2}{c}{$<$ 141} & 1.39 & $\ddagger$ & \multicolumn{2}{c}{$<$ 223} & 
1.44 & $\ddagger$,\citetalias{malesani12} & $<$2.5 \\ 
\hline
\multicolumn{12}{c}{Other hosts}\\
\hline
\object{050915A} & 2.527 & \citetalias{jakobsson12,kruhler12b} & 
\multicolumn{2}{c}{$<$ 59} & 1.39 & $\ddagger$ & \multicolumn{2}{c}{$<$ 1204} & 
9.48 & $\ddagger$,\citetalias{malesani12} & $<$2.4 \\ 
\dots & \dots & \dots & \multicolumn{2}{c}{$<$ 44} & 5.50 & $\ddagger$ & 
\multicolumn{2}{c}{$<$ 2521} & 9.48 & $\ddagger$,\citetalias{malesani12} & 
$<$2.7 \\ 
\dots & \dots & \dots & \multicolumn{2}{c}{$<$ 37} & 9.00 & $\ddagger$ & 
\multicolumn{2}{c}{$<$ 3032} & 9.48 & $\ddagger$,\citetalias{malesani12} & 
$<$2.8 \\ 
\object{060505} & 0.0889 & \citetalias{ofek06gcn} & \multicolumn{2}{c}{$<$ 37}
 & 5.50 & $\ddagger$ & \multicolumn{2}{c}{$<$ 1.50} & 1.90 & 
\citetalias{castroceron10} & $\sim0$ \\ 
\dots & \dots & \dots & \multicolumn{2}{c}{$<$ 52} & 9.00 & $\ddagger$ & 
\multicolumn{2}{c}{$<$ 2.53} & 1.90 & \citetalias{castroceron10} & $<$0.1 \\ 
\dots & \dots & \dots & \multicolumn{2}{c}{$<$ 63} & 1.43 & $\ddagger$ & 
\multicolumn{2}{c}{$<$ 1.05} & 1.90 & \citetalias{castroceron10} & $\sim0$ \\ 
\object{060814} & 1.92 & \citetalias{jakobsson12,kruhler12b,salvaterra12} & 
\multicolumn{2}{c}{$<$ 430} & 1.43 & $\ddagger$ & \multicolumn{2}{c}{$<$ 4823}
 & 31.20 & $\ddagger$,\citetalias{malesani12} & $<$2.5 \\ 
\object{070808} & unknown & \dots & \multicolumn{2}{c}{$<$ 156} & 1.43 & 
$\ddagger$ & \multicolumn{2}{c}{\dots} & \dots & \dots & \dots \\ 
\hline
\end{tabular}
\tablefoot{
The horizontal lines divide the GRB-SN and the $z<1$ TOUGH unbiased subsets (see Section~\ref{sec:sample}) and the hosts which do not belong to any of these subsets. GRBs \object{050525A} and \object{060218}  belong to the first two subsets. For non-detected targets $3\sigma$ limits are reported.
}
\tablefoottext{a}{Assuming radio spectral index $\alpha=-0.75$ and applying the calibration of \citet{bell03}.}
\tablefoottext{b}{From UV continuum unless noted otherwise. Not corrected for dust attenuation.}
\tablefoottext{c}{The symbol $\ddagger$ indicates that we derived the SFR from fluxes reported in the reference using the calibration of \citet{kennicutt} or \citet{savaglio09}.}
\tablefoottext{d}{Visual extinction calculated from the ultraviolet extinction $A_{\rm UV}=2.5\log(\mbox{SFR}_{\rm radio}/\mbox{SFR}_{\rm UV})$ assuming an SMC extinction curve, which gives $A_{\rm V}=A_{\rm UV}/2.2$.}
\tablefoottext{e}{From the [\ion{O}{2}] line.}
\tablerefs{$\ddagger$: This work,
 \GiveRef{bloom01}, 
 \GiveRef{tinney98}, 
 \GiveRef{galama99gcn}, 
 \GiveRef{hjorth00}, 
 \GiveRef{castrotirado01}, 
 \GiveRef{price02000911}, 
 \GiveRef{price02}, 
 \GiveRef{infante01gcn}, 
 \GiveRef{price03}, 
 \GiveRef{soderberg04b}, 
 \GiveRef{vreeswijk06}, 
 \GiveRef{hjorthnature}, 
 \GiveRef{prochaska04}, 
 \GiveRef{soderberg06}, 
 \GiveRef{cenko05gcn}, 
 \GiveRef{soderberg07}, 
 \GiveRef{foley05gcn}, 
 \GiveRef{fynbo05gcn}, 
 \GiveRef{sollerman07}, 
 \GiveRef{soderberg05gcn2}, 
 \GiveRef{jakobsson12}, 
 \GiveRef{pian06}, 
 \GiveRef{price06gcn}, 
 \GiveRef{thone06gcn}, 
 \GiveRef{fynbo09}, 
 \GiveRef{jaunsen07gcn}, 
 \GiveRef{kruhler12b}, 
 \GiveRef{ofek06gcn}, 
 \GiveRef{salvaterra12}, 
 \GiveRef{frail03}, 
 \GiveRef{michalowski09}, 
 \GiveRef{vreeswijk01radio}, 
 \GiveRef{stanway10}, 
 \GiveRef{berger}, 
 \GiveRef{berger03}, 
 \GiveRef{hatsukade12}, 
 \GiveRef{fox03}, 
 \GiveRef{vanderhorst05}, 
 \GiveRef{watson11}, 
 \GiveRef{wiersema08}, 
 \GiveRef{soderberg04gcn}, 
 \GiveRef{castroceron10}, 
 \GiveRef{christensen04}, 
 \GiveRef{savaglio09}, 
 \GiveRef{svensson10}, 
 \GiveRef{ovaldsen07}, 
 \GiveRef{cano11}, 
 \GiveRef{malesani12}.
}
\end{table*}

\begin{figure*}[t!]
\begin{center}
\includegraphics[width=\textwidth]{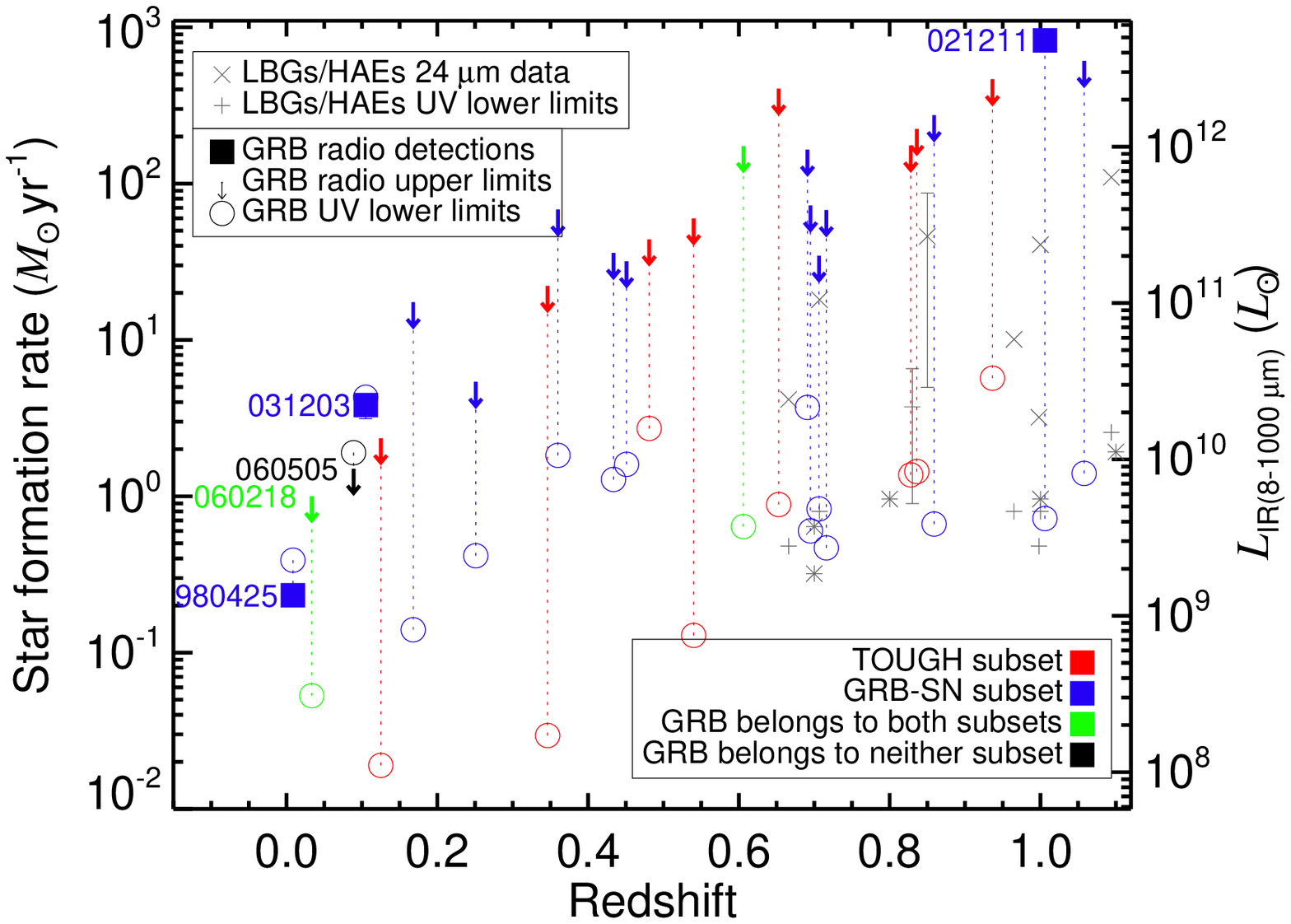}
\end{center}
\caption{Star formation rates (SFRs) as a function of redshift of GRB host galaxies. {\it Squares} and {\it arrows} denote SFRs  derived from radio detections and $3\sigma$ upper limits, respectively. {\it Circles} denote lower limits on SFRs derived from the ultraviolet (UV) data. 
For a given GRB, the radio and UV SFRs are connected by a {\it dotted line}. GRBs are color-coded depending on whether they belong to the $z<1$ TOUGH unbiased subset ({\it red symbols}), the GRB-SN subset ({\it blue symbols}), both ({\it green symbols}), or none ({\it black symbols}). 
The right $y$-axis gives the corresponding infrared luminosity according to SFR$(\msunyr)=1.72\times10^{-10}L_{\rm IR} (L_\odot)$ \citep{kennicutt}.
The three low-redshift hosts (GRB 980425, 031203, and 060505) are consistent with no dust attenuation as their SFR$_{\rm radio}$ are similar to SFR$_{\rm UV}$. On the other hand, huge dust attenuation must be invoked to explain a very high SFR$_{\rm radio}$ of the host of GRB 021211. 
 {\it Crosses} and {\it plus symbols} indicate the $24\,\mu$m and UV SFRs of Lyman break galaxies \citep[LBGs;][]{basuzych11} and H$\alpha$ emitters \citep[HAEs; mean values with standard deviations are shown;][]{sobral09}. GRB hosts are consistent with these populations.
}
\label{fig:sfrz}
\end{figure*}

Our photometry measurements are presented in Table~\ref{tab:flux}.  Three (GRB 980425, 021211, 031203) out of twenty targeted hosts were detected (not counting upper limits from the literature). 
 Two out of these detections are in fact the first- and third-closest GRBs in our sample. None of the hosts in the TOUGH subset has been detected.
Hence, this program (\citealt{michalowski09}, \citealt{watson11}, and this paper) increases the number of the radio-detected GRB hosts from two (GRB 980703, \citealt{bergerkulkarni}; GRB 000418, \citealt{berger}) to five. Recently the host of GRB 031203 has also been detected at $5.5$ GHz by \citet{stanway10}.

We assume that the entire flux is due to star formation and not active galactic nucleus (AGN) activity, which is a well-tested hypothesis for GRB hosts  \citep[see discussion in][]{michalowski08,watson11}. 

The SFRs derived from our radio data as well as from the ultraviolet (UV) data are presented in Table~\ref{tab:flux} and are shown  as a function of redshift on Figure~\ref{fig:sfrz}.
The radio SFRs (SFR$_{\rm radio}$) were calculated from the empirical formula of \citet[][see Section~4.2 of \citealt{michalowski09} for discussion of its applicability to GRB hosts]{bell03}  assuming a radio spectral index\footnote{Defined as $F_\nu \propto \nu^\alpha$, i.e.~$\alpha_{\nu_1}^{\nu_2}= \log [F_\nu(\nu_2)/F_\nu(\nu_1)] / \log (\nu_2/\nu_1)$.} $\alpha=-0.75$ \citep{condon,ibar10}.  This choice of spectral index  has relatively small impact on derived SFRs, because our observed $1.4$ GHz data probe close to the rest-frame $1.4$ GHz, at which the flux--SFR conversion is calibrated. Namely if we assumed a flat index $\alpha=0$, then we would obtain SFRs  $\sim25$--$40$\% lower at $z=0.5$--$1$. On the other hand, if we assumed a steeper value $\alpha=-1$ (or $-1.5$), then we would obtain SFRs $\sim10$--$20$\% ($\sim35$--$70$\%) higher at  $z=0.5$--$1$. 

The limit on the SFR of the GRB 980425 host  based on $8.64$ GHz data is not consistent with the value derived from the $4.80$ GHz data, because, for consistency, a spectral slope of $\alpha=-0.75$ was assumed, whereas in reality it is steeper (see \citealt{michalowski09} and section~\ref{sec:alpha}).

In order to assess amount of the dust attenuation in GRB hosts we compared their SFRs derived from the UV emission (SFR$_{\rm UV}$)  with SFR$_{\rm radio}$. In Table~\ref{tab:flux} we compiled the SFR$_{\rm UV}$ (mostly from $0.28\,\mu$m continuum data) from the literature \citep{castroceron10,savaglio09,christensen04, jakobsson12,ovaldsen07, svensson10}. The de-reddened SFRs given by \citet{savaglio09} were reddened based on their reported $A_V$.
For the hosts of GRB 051016B, 051117B, 060814 and 070318  
we calculated the SFR$_{\rm UV}$ from $V$-, $B$-, and $R$-band fluxes, respectively, reported by \citet{ovaldsen07} 
and \citet{malesani12},
 which correspond to the rest-frame UV emission at the redshifts of the hosts.
For the hosts of  GRB 040924 and 061021 there are no UV continuum data available, so 
we calculated  SFR$_{[\mbox{\ion{O}{2}}]}$
from the flux reported by \citet{wiersema08} and \citet{jakobsson12}, respectively, applying the conversion of 
 \citet[][their equation~(4)]{kewley04}.

We assume that SFR$_{\rm radio}$ reflects the total amount of star formation in GRB hosts. 
Hence, an approximate estimate of the dust attenuation in the ultraviolet may be obtained by dividing the radio SFR and SFR$_{\rm UV}$: 
\begin{equation}
A_{\rm UV}=2.5\log\frac{\mbox{SFR}_{\rm radio}}{\mbox{SFR}_{\rm UV}} \, \mbox{mag}
\end{equation}
 The resulting attenuations are presented in Table~\ref{tab:flux}.
The uncertainties of SFR$_{\rm radio}$ and SFR$_{\rm UV}$ are of the order of a factor of two \citep{bell03,kennicutt}, so the uncertainties of the $A_{\rm UV}$ estimates are of the order of a factor of $2\sqrt{2}\sim2.8$ ($\sim1.1$ mag).

\section{Discussion}
\label{sec:discussion}

\subsection{The ULIRG Nature of the Host of GRB 021211}

Our $1.43$ GHz detection of the host of GRB 021211 ($\sim10\sigma$)  corresponds to SFR of $\sim825\, \msunyr$ (Table~\ref{tab:flux}), which places it in the category of ultraluminous infrared galaxies (ULIRGs; $L_{\rm IR}>10^{12}\,L_\odot$, or SFR$\mbox{}\gtrsim172\,\msunyr$ using the conversion of \citealt{kennicutt}). This is the highest SFR ever reported for a GRB host \citep[compare with][]{berger, bergerkulkarni,michalowski08,stanway10,watson11}.

 Because of this unusually high SFR
 we present the  investigation of  the data quality for this object.  
 We verified that the source is not due to an uncleaned bright source nearby. Moreover,  the astrometry of our VLA map and the VLT image \citep{dellavalle03} are consistent. Namely, for three radio sources that are detected in the optical image\footnote{With the following radio R.A. and decl.: 08:09:01.315, +06:43:03.91; 08:09:12.842, +06:43:54.53; and 08:09:11.746, +06:44:05.87.} we measured the mean offsets between the radio and optical positions consistent with zero: $\Delta\alpha=-0.27''\pm0.32''$ and $\Delta\delta=0.14''\pm0.39''$. Finally, the median ratio of fluxes of bright objects detected in our radio map to the fluxes reported in the NRAO VLA Sky Survey  \citep[NVSS;][]{nvss} is  $\sim1.19^{+0.31}_{-0.36}$, confirming the accuracy of the flux calibration.

However, our detection is
inconsistent with a non-detection at $2.1$ GHz presented by \citet[][$S_{2.1}<34.2\,\mu$Jy at $3\sigma$]{hatsukade12} as it implies an extremely steep (and unphysical) spectral index $\alpha^{2.10}_{1.43}<-5.9$. 
Extrapolating from our $1.43$~GHz detection, the expected signal at $2.1$~GHz would be $\sim250\,\mu$Jy assuming $\alpha=-0.75$.  However the equatorial declination of GRB 021211 of $+6^\circ44'$ makes it difficult to observe it using east--west arrays, such as ATCA, as the beam is highly elongated, i.e.~$2'' \times 51''$ at $2.1$ GHz. 
This may pose some problems in the detection of sources, and indeed in our VLA map we have found two additional sources\footnote{With the following VLA R.A. and decl.: 08:09:10.992, +06:41:26.20; and 08:09:31.704, +06:41:58.00.}
 with $1.43$ GHz fluxes of $\sim650$ and $350\,\mu$Jy, respectively, which are not detected in the $2.1$ GHz ATCA map of \citet{hatsukade12}. Further observations at various radio  frequencies are needed to resolve this issue.  

Our $1.43$ GHz detection is consistent with the (sub)millimeter limits of \citet{smith05} and \citet{priddey06}. Namely,
they did not detect the host of GRB 021211 at $850\,\mu$m ($0.3\pm1.9$ mJy) and $1.2$ mm ($0.07\pm0.53$ mJy), respectively. The  $850\,\mu$m limit implies a submillimeter-to-radio spectral index $\alpha^{350}_{1.4}<0.53$, consistent with most of the models presented by \citet[][their Figures~1 and 3, respectively]{carilli99, carilli01b}.
Assuming spectral energy distribution (SED) templates of Arp 220 and M82 \citep{silva98}, the $1.2$ mm limit corresponds to a $3\sigma$ limit of SFR$\mbox{}\lesssim700$--$1100\,\msunyr$ (for the $850\,\mu$m limit these estimates are $\sim30$\% higher), consistent with our radio SFR$\mbox{}\sim825\,\msunyr$ (Table~\ref{tab:flux}).
However, this $1.2$ mm flux limit corresponds to an SFR$\mbox{}\sim100$--$500\,\msunyr$ for many other SED templates \citep{silva98,iglesias07,michalowski08,michalowski10smg}. Hence, deeper (sub)millimeter observations (rms$\mbox{}\sim0.1$--$0.2$ mJy) are needed to verify if our radio detection is inconsistent with  (sub)millimeter data, which would indicate a significant AGN contribution to the radio flux of the host of GRB 021211.

\subsection{Star Formation Rates of the GRB Host Population}

\begin{figure*}[t!]
\begin{center}
\includegraphics[width=\textwidth]{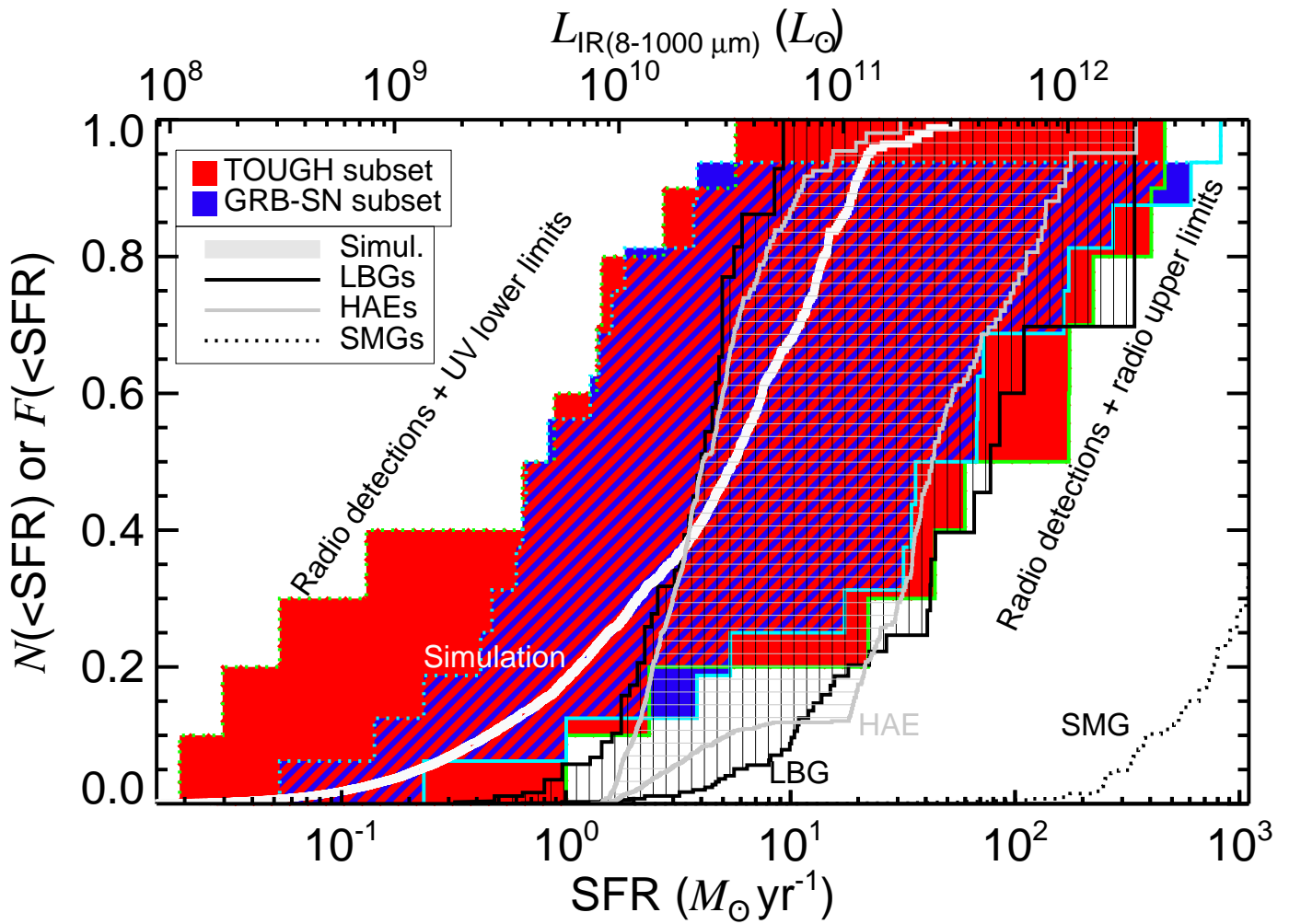}
\end{center}
\renewcommand{\baselinestretch}{1.}
\caption{Cumulative distribution of SFRs of GRB hosts in the $z<1$ TOUGH unbiased ({\it red area}) and the GRB-SN ({\it  blue area}) subsets. The high-SFR boundaries ({\it colored solid lines}) are constructed using the detections and limits of SFR$_{\rm radio}$ (Table~\ref{tab:flux}), whereas the low-SFR boundaries ({\it colored dotted lines}) are constructed using the SFR$_{\rm UV}$ for galaxies not detected in the radio. 
The upper $x$-axis gives the corresponding infrared luminosity according to SFR$(\msunyr)=1.72\times10^{-10}L_{\rm IR} (L_\odot)$ \citep{kennicutt}.
We found that at least $\sim63$\% ($\ge$15/24) of GRB hosts at $z\lesssim1$ have SFR$\mbox{}<100\, \msunyr$ and only $\lesssim8$\% ($\le$2/24) could have SFR$\mbox{}>500\,\msunyr$.
For comparison, the SFR distributions of $z=0.51$ simulated galaxies \citep{croton06}, $z\sim1$ Lyman break galaxies  \citep[LBG;][]{basuzych11}, $z\sim0.84$ H$\alpha$ emitters \citep[HAE;][]{sobral09}, and $z\sim2$--$3$ submillimeter galaxies  \citep[SMGs;][]{michalowski10smg,michalowski10smg4} are shown ({\it labelled lines}, of which the right lines represent dust-corrected SFRs).  These distributions were weighted by SFR (so they reflect the fraction of total star formation in the sample contributed by galaxies with SFRs lower than a given SFR) to allow a comparison with the GRB host population, which is likely selected based on SFRs (see Section~\ref{sec:unb}). It is evident that current SFR limits imply that the GRB host population is consistent with star-forming galaxies at similar redshifts (simulated, LBGs, and HAEs) and is inconsistent with SMGs. 
Since we did not detect most of the targets, the distributions of the $z<1$ TOUGH unbiased and GRB-SN subsets are consistent (the overlap of the blue and red areas is significant).
}
\label{fig:cumulSFR}
\end{figure*}

 The SFRs of GRB hosts are
shown as a cumulative distribution 
on Figure~\ref{fig:cumulSFR}. The high-SFR boundaries were calculated using the radio detections and upper limits, whereas the low-SFR boundaries were obtained by substituting the radio SFR upper limits with the lower limits  from the UV. We found that at least $\sim63$\% ($\ge$15/24)\footnote{$\ge50$\% ($\ge$5/10) for TOUGH subset only.} of all our GRB hosts at $z\lesssim1$ have SFR$\mbox{}<100\,\msunyr$ and only $\lesssim8$\% ($\le$2/24)\footnote{$0$\% (0/10) for TOUGH subset only.} could have SFR$\mbox{}>500\,\msunyr$. This implies that it is rare ($\lesssim33$\% chance, $\le$8/24)\footnote{$\le50$\% ($\le$5/10) for TOUGH subset only.} for a GRB to reside in an ULIRG. 
 This is consistent with the contribution of ULIRGs to the cosmic star formation history being $<10\%$ at $z<1$ \citep{lefloch05}.

Even though high star-forming GRB hosts are rare at $z\lesssim1$, the SFR of GRB 021211 alone constitutes as much as $\sim22$\% of the summed SFR of all $z\lesssim1$ GRB hosts, even when we sum over radio upper limits (and hence its contribution is higher in reality).  Hence, such high star-forming GRB hosts likely dominate the contribution of this population to the cosmic star formation history.

 The average radio SFR of GRB hosts can be assessed using the average radio flux of the GRB hosts undetected in our radio observations. For each host we converted the flux at the GRB position at the observed frequency to that at the rest-frame $1.43$ GHz, using a radio spectral index of $-0.75$.
In this way we obtained  a weighted mean of the flux  equal to $ -13 \pm16 \,\mu$Jy. 
Hence, we did not detect the GRB host population even when  averaging the data. 
At least such level of rms has to be reached in future GRB host surveys to obtain significant number of detections.
At the mean redshifts of these hosts, $z=0.53$, this corresponds to a $3\sigma$ upper limit of  SFR$_{\rm radio}< 15\,\msunyr$. Hence, the general population of GRB hosts is below the LIRG limit ($L_{\rm IR}<10^{11}\,L_\odot$, or SFR$\mbox{}\lesssim17.2\,\msunyr$ using the conversion of \citealt{kennicutt}). 
It is expected that LIRGs do not dominate our GRB host sample, because
 LIRGs dominate the cosmic star formation history only above $z\sim0.7$ 
 (with their contribution rising to $\sim70$\% at $z=1$, \citealt{lefloch05}; and staying at this level at least up to $z\sim2.3$, \citealt{magnelli11}).

The full ALMA with 50 antennas will reach an rms sensitivity of $\sim0.023$\,mJy at 345 GHz in 1 hr\footnote{Assuming fourth octile of  water vapor; \newline \urltt{http://almascience.eso.org/call-for-proposals/sensitivity-calculator}}. This corresponds to SFR$\mbox{}\sim5$--$20\,\msunyr$ at $z=10$  \citep[using SED templates of][]{silva98,iglesias07,michalowski08,michalowski10smg}, so ALMA will easily detect GRB hosts basically at any redshift within a few hours, because the UV lower limits on SFRs are of the order of $\sim1\,\msunyr$ (Table~\ref{tab:flux}).

To summarize, the overall picture is that $z\lesssim1$ GRB hosts have modest SFRs \citep[as suggested by][]{stanway10}, but a small fraction ($\sim4$--$8$\%) of them 
have undergone an extreme star formation episode. However the latter claim suffers from poor number statistics.

\subsection{The Relation to Other Galaxies: Do GRBs Trace Star Formation in an Unbiased Way?}
\label{sec:unb}

 In order to investigate whether the GRB host population is consistent with the general population of star-forming galaxies at similar redshifts, we show their SFR distribution on Figure~\ref{fig:cumulSFR}. The comparison to other galaxies must be done carefully, because the probability that a galaxy with given SFR is included in a usual galaxy sample depends only on the number density of such objects (as long as this SFR corresponds to a flux higher than the sample selection threshold). This is not the case for a GRB host sample,  because, assuming that GRBs trace star formation in an unbiased way, a galaxy with higher SFR is more likely to host a GRB and, in turn, to be selected into the GRB host sample \citep[e.g.][]{natarajan97,fynbo01b}. In order to account for this, we weighted the cumulative distributions of other galaxies by their SFRs, i.e.~the curves for other galaxies correspond to the fraction of total star formation in the sample contributed by galaxies with SFRs lower than a given SFR.

It is apparent from Figure~\ref{fig:cumulSFR} that the SFR distributions of GRB hosts and of simulated galaxies at $z=0.51$ \citep{croton06}\footnote{\urltt{http://tao.it.swin.edu.au/mock-galaxy-factory/}}, produced in a semi-analytical model and based on the Millenium simulation \citep{springel05}, are fully consistent. 

Similarly, the SFR distribution of $z\lesssim1$ GRB hosts is consistent with that of $z\sim1$ Lyman break galaxies \citep[LBGs; from][SFRs from  $24\,\mu$m and rest-frame UV photometry]{basuzych11} and $z\sim0.84$ H$\alpha$ emitters \citep[HAEs; from][SFRs from $24\,\mu$m photometry and H$\alpha$ fluxes]{sobral09}. We note that the median stellar masses of these LBGs  \citep[$M_*\sim10^{9.5}\,\msun$;][]{basuzych11} and HAEs \citep[$M_*\sim10^{10.1}\,\msun$;][]{sobral11}  are also consistent with that of GRB hosts \citep[$M_*\sim10^{9.3-9.7}\,\msun$;][]{castroceron10,savaglio09}.  Moreover, as shown in Section~\ref{sec:a}, the dust attenuation we derived for GRB hosts is consistent with that of LBGs and HAEs.

An apparent inconsistency  at low SFRs of the GRB host samples with the LBG and HAE populations (the GRB population extends to lower SFRs) is an effect of higher flux detection threshold for the latter. Namely, the limiting magnitude for LBGs of $u < 24.5$ mag corresponds to SFR$\mbox{}>0.5\,\msunyr$, whereas the limiting luminosity for HAE of $L_{\rm H\alpha} > 10^{41.5}$ erg s$^{-1}$ corresponds to SFR$\mbox{}>2.5\,\msunyr$. Hence,  galaxies with SFR$\mbox{}<0.5\,\msunyr$ are not present in the LBG and HAE samples, because they are below the detection limits. Indeed when we restricted the simulated galaxies (which are consistent with the GRB host population) to galaxies above these limits, their distributions are consistent with those of LBGs and HAEs.
 
A Kolmogorov--Smirnov (K-S) test resulted in a probability of $\sim15$\% that the UV SFRs of  all our GRB hosts and $z<1$ LBGs are drawn from the same population. However, for $z\sim0.84$ HAEs such probability is negligible ($\sim10^{-11}$), showing that  HAEs have systematically higher UV SFRs than GRB hosts.  Figures~\ref{fig:sfrz} and \ref{fig:cumulSFR} show that our current limits on the radio SFRs of GRB hosts are not deep enough to test whether the total SFRs of HAEs are also higher than that of GRB hosts.

As shown in Figure~\ref{fig:cumulSFR}, the SFRs of the $z\lesssim1$ GRB hosts are clearly inconsistent with those of submillimeter galaxies (SMGs), dusty high star-forming $z\sim2$--$3$ galaxies \citep[][SFRs from total infrared emission and rest-frame UV photometry]{michalowski10smg,michalowski10smg4}, even when only radio upper limits for GRB hosts are taken into account (colored solid lines).
This is also expected from the fact that GRB hosts are much less massive \citep[$M_*\sim 10^{9.3-9.7}\,\msun$;][]{castroceron10,savaglio09} than SMGs \citep[$M_*\sim10^{10.4-11.3}\,\msun$;][]{borys05,michalowski10smg,michalowski12mass,hainline11,bussmann12,yun12}.
 We note however, that the stellar mass estimates for GRB hosts are based on samples biased against dusty galaxies, so the general population of GRB hosts may include galaxies as massive as SMGs.

The inconsistency of the $z\lesssim1$ GRB host and SMG populations is also revealed by the K-S test giving a negligible probability of $\sim10^{-8}$ that the UV SFRs of GRB hosts and SMGs are drawn from the same population. If we restrict the analysis to $z<1$ SMGs then the probability increases to $\sim63$\% (consistent samples), but these $z<1$ SMGs have median SFR$_{\rm IR}\sim60\,\msunyr$, very close to our limits of SFR$_{\rm radio}$ for GRB hosts, so if GRB hosts are similar to $z<1$ SMGs, then the majority of them would need to be just below the detection limits,  which is unlikely.

We note that the comparison between GRB hosts, SMGs, LBGs, and HAEs involves SFRs derived from the radio, $24\,\mu$m, total infrared, UV, and H$\alpha$ luminosities, but we do not expect any systematic offset between these estimates. Namely, \citet[][their Figure~2]{elbaz10} showed that the observed $24\,\mu$m luminosity is correlated with the total infrared luminosity, which, in turn, is correlated with the radio luminosity \citep{condon}. Similarly, \citet[their Figure~8]{wijesinghe11} showed that SFRs from H$\alpha$ and UV are well correlated.

To summarize, our data  allow a significant range of SFRs for  $z\lesssim1$ GRB hosts, so their distribution
is consistent with that of the general population of galaxies (at least with LBGs and simulated galaxies) having a median SFR of a few \msunyr. If this conclusion is confirmed  in deeper radio observations, 
it would indicate that GRBs trace a large fraction of the overall star formation, and are therefore less biased indicators than once thought.
On the other hand, if deeper radio observations reveal that the total SFRs of GRB hosts are very close to their UV SFRs, then GRB hosts will not be consistent with tracing the overall population of star-forming galaxies.

Indeed,
 there are some indications that GRB hosts are different than other galaxies at similar redshifts.  Previous studies have showed that GRB hosts are metal-poor \citep{fynbo03,prochaska04,gorosabel05,sollerman05,hammer06,stanek06,wiersema07,christensen08,modjaz08,modjaz11,levesque10c,levesque10,leloudas11}, fainter, and more compact than SN hosts, as well as that GRBs themselves are much more concentrated in the UV-bright regions of their hosts than SNe  \citep{fruchter06,bloom02}. Moreover, \citet[][their Figures~3 and 6]{wainwright07} found that in the $z<1$ GRB host population the ratio of irregular to regular galaxies is 2:1, different than 1:2 for field galaxies; and that GRB hosts are a factor of two smaller than field galaxies. This suggests a potential bias in the GRB  host population (though all these studies were based on optically-biased GRB host samples). 
 
 However, \citet[][their Figure~3]{leloudas10} showed that the distribution of GRBs within their hosts is in fact consistent with that of type-Ic SNe and of Wolf--Rayet stars, whereas \citet[][their Table~3 and Figure~8]{svensson10} did not find any significant difference between the absolute magnitudes of GRB and SN hosts. 
 Moreover, \citet[][their Figure~2]{fynbo06} and \citet[][their FIgure~18]{savaglio09} showed that the metalicity of GRB hosts is consistent with \citep[or even higher than; see also][]{savaglio12} that of damped Ly$\alpha$ systems at corresponding redshifts.
  Similarly, \citet{levesque10b} found that some GRBs explode in metal-rich environments. Moreover, optically-dark GRBs, missed in most previous studies, were claimed to be hosted in more dusty and metal-rich hosts \citep{perley09,greiner11,hunt11,kruhler11,svensson12}. 
 Finally, \citet{conselice05} found that GRB hosts are smaller than field galaxies, but only at $z<1.2$; and that these two populations at $z<1$ are consistent with regards to their concentrations and asymmetries. Hence, the biases of the GRB host sample in terms of morphology and metallicity are far from being well understood.

 Regarding observational biases, \citet{fynbo09} showed that the GRB host sample selected based on a requirement of an optical spectroscopic redshift is not representative for all GRBs, likely biased against dusty objects. Moreover, SNe exploding in (U)LIRGs are expected to be highly extinguished by dust \citep{mattila04,mattila07,mattila12,melinder12}, implying that samples of optically-selected GRBs may miss completely those exploding in high star-forming galaxies. 
However, Figure~\ref{fig:cumulSFR} does not reveal any differences between the  radio properties of the optically biased pre-{\it Swift} sample of GRB-SN hosts and those of the $z<1$ TOUGH unbiased subset. Namely, the high-SFR bounds of the distributions are similar. This is mainly because we did not detect the majority of the targets and the limiting depths were similar. 
Similarly, the K-S test resulted in $40$\% probability that the SFR$_{\rm UV}$ for the TOUGH and GRB-SN subsets are drawn from the same parent population.
Larger samples and deeper radio data are necessary to investigate this issue.

\subsection{Ultraviolet Attenuation}
\label{sec:a}

\begin{figure*}[t]
\begin{center}
\includegraphics[width=\textwidth]{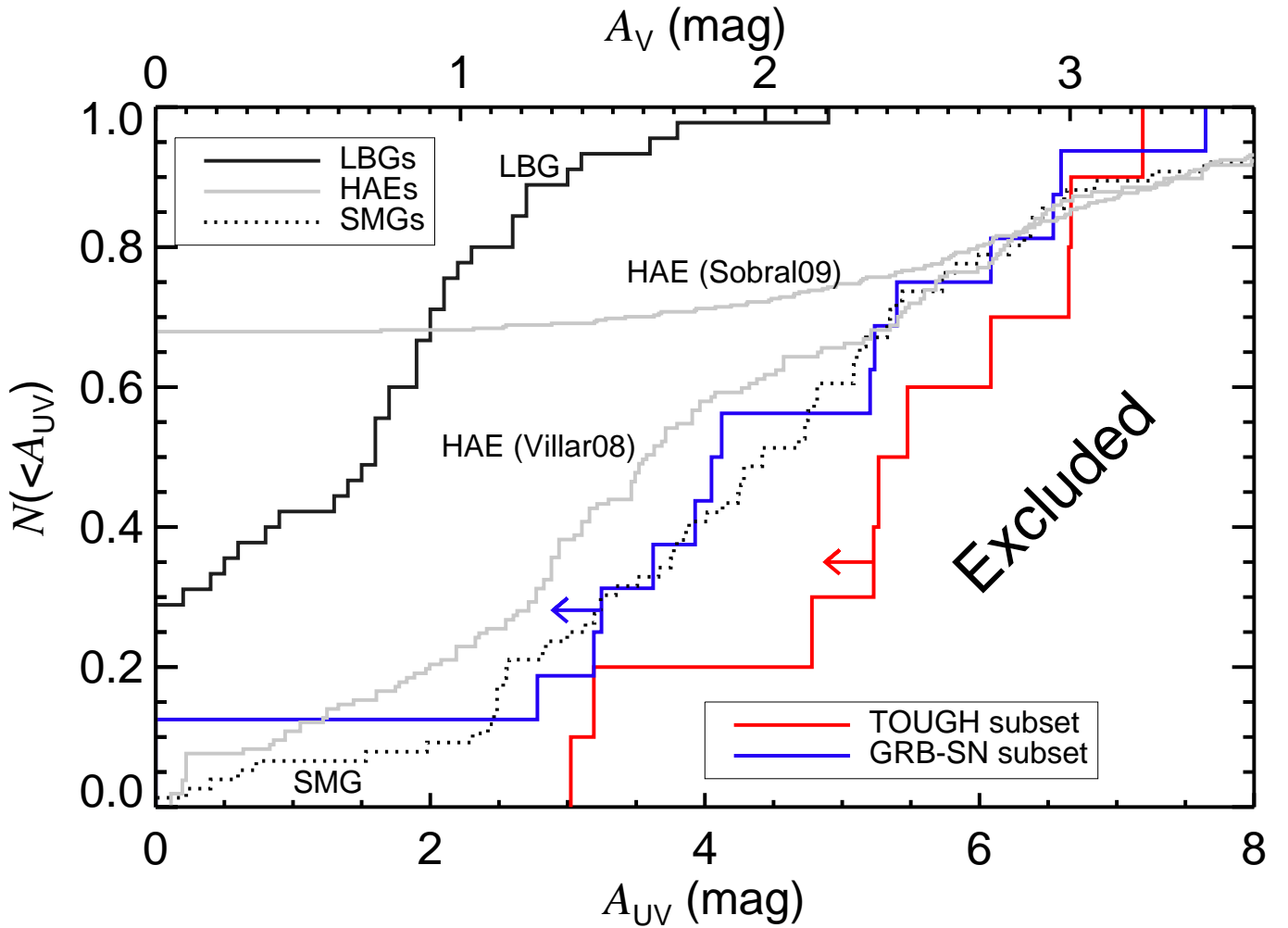}
\end{center}
\caption{Cumulative distribution of ultraviolet (and optical) dust attenuation of GRB hosts in the $z<1$ TOUGH unbiased ({\it red solid line}) and GRB-SN ({\it blue solid line}) subsets. Since the distributions are mostly based on radio non-detections they should be taken as upper limits to $A_{\rm UV}$, i.e.~the area to the right of the lines is ruled out by the data.
The upper $x$-axis  was derived by assuming an SMC extinction curve \citep[$A_{\rm UV}=2.22\, A_V$;][]{gordon03}. 
We found that $\gtrsim88$\% ($\ge$21/24) of the GRB hosts have $A_{\rm UV}<6.7$ mag, i.e.~$A_V<3$ mag.
For comparison, the distributions of $z\sim1$ Lyman break galaxies  \citep[LBG;][]{basuzych11}, $z\sim0.84$ H$\alpha$ emitters \citep[HAE;][]{villar08,sobral09}, and $z\sim2$--$3$ submillimeter galaxies  \citep[SMGs;][]{michalowski10smg,michalowski10smg4} are shown ({\it labelled black lines}). The attenuation at H$\alpha$ was converted to that in the ultraviolet by assuming an SMC extinction curve \citep[$A_{\rm UV}=1.78\,A_{{\rm H}\alpha}$;][]{gordon03}.  The distributions for GRB hosts are consistent with those of LBGs and HAEs,  yet likely inconsistent with that of SMGs.
}
\label{fig:cumulAV}
\end{figure*}

The cumulative distribution of the dust attenuation is shown on Figure~\ref{fig:cumulAV}. It should be seen as an upper limit, because it is based mostly on radio non-detections.
We found that $\gtrsim88$\% ($\ge$21/24)\footnote{$\ge80$\% ($\ge$8/10) for TOUGH subset only.} GRB hosts at $z\lesssim1$ have $A_{\rm UV}<6.7$ mag, i.e.~$A_V<3$ mag assuming a Small Magellanic Cloud (SMC) extinction curve \citep[$A_{\rm UV}=2.22\, A_V$;][]{gordon03}. This is consistent with the previous observational  \citep{kann06,schady07,schady10,schady12,kann10,savaglio09,han10,liang10, greiner11,watson11b,watson12,zafar10,zafar11} and theoretical \citep{lapi08,mao10} results that GRB hosts are weakly obscured by dust with very few exceptions \citep{tanvir,tanvir08,castroceron06,lefloch,michalowski08,kupcuyoldas10,hunt11,kruhler11,svensson12}.
This result is also confirmed for the entire TOUGH sample \citep{hjorth12}, which is {\it not} biased against dusty systems. 
Our finding is also consistent with a fraction of $5$--$36$\% of core collapse SNe in normal galaxies with $A_V>3.7$ mag \citep[Section 2.4 of][]{mattila12}.

It has been claimed that dust is responsible for the optical faintness of the so-called dark GRBs, i.e.~those with optical--to--X-ray spectral index $\beta_{\rm OX}<0.5$ \citep{ramirezruiz02,jakobsson04,perley09,greiner11,kruhler11}.  None of our $z\lesssim1$ GRBs were dark \citep[though there is no X-ray data for four  pre-{\it Swift} GRBs in our sample]{jakobsson04,butler05,fynbo09,xu09}. Hence, none of the afterglows of the GRBs in our $z\lesssim1$ sample seems to be particularly dust-obscured  (though the effect of dust at $z<1$ is less severe than at higher redshifts, i.e.~for the same amount of dust a GRB at higher redshift may be classified as ``dark'', because the observed-frame optical emission corresponds to the UV wavelengths strongly affected by dust). This is consistent with low levels of dust attenuation we find for their hosts, but we note that, in principle, the spatially-integrated attenuation of a host reported here may be inconsistent with the line-of-sight extinction derived from an afterglow, as the latter probes dust distributed only over a narrow opening angle. 

GRB 021211, with a radio-detected host, was initially called ``dark'' due to its optical faintness \citep{crew03,fox03,li03}, but the lack of X-ray data makes it impossible to classify it according to the more rigorous $\beta_{\rm OX}$ definition. However, \citet{fox03} found that the optical afterglow was not severely reddened, so the GRB must have occurred away from the highly dust-attenuated regions suggested by our radio detection.

We also note that only $\lesssim50$\% of  all our GRB hosts at $z\lesssim1$ could have $A_V>2$ mag ($A_{\rm UV}>4.4$ mag),  i.e.~similar dust attenuation levels to that in SMGs \citep{smail04,swinbank04,borys05,michalowski10smg,michalowski10smg4,hainline11}. Indeed, in Figure~\ref{fig:cumulAV} we show that the $A_{\rm UV}$ distribution for SMGs \citep[from][]{michalowski10smg,michalowski10smg4} displays high attenuation levels,  very likely inconsistent with that of $z\lesssim1$ GRB hosts, given that the lines for GRB hosts represent upper limits on dust attenuation. This is  consistent with a tendency that the presence of a GRB typically selects dwarf galaxies that are generally less dusty. However, the most star-forming GRB hosts may contain significant amounts of dust, comparable to those of SMGs \citep[as suggested by][]{michalowski08}.

In Figure~\ref{fig:cumulAV} we also show the attenuation distribution of $z\sim1$ LBGs from \citet{basuzych11}. Our GRB host distribution is mostly based on upper limits on $A_{\rm UV}$, but it is clear that these two distribution may be consistent, i.e.~the fraction of objects with very low dust attenuation is similar ($\sim20$--$30$\%) and the LBG distribution is always below that of the GRB hosts. Deeper radio or far-IR data would be necessary to confirm that these samples are truly consistent. If this is the case, then we can expect a median $A_{\rm UV}\sim1.6$ mag for GRB hosts and therefore that their SFR$_{\rm radio}$ should be, on average, a factor of $\sim4$ higher than their SFR$_{\rm UV}$. 
Assuming a typical SFR$_{\rm UV}=1\,\msunyr$, then the SFR$_{\rm radio}=4\,\msunyr$ corresponds to the observed-frame $1.43$ GHz fluxes of $33$, $8$, and $3\,\mu$Jy at $z=0.3$, $0.6$, and $1.0$, respectively. Hence, if the attenuation of GRB hosts is indeed similar to that in LBGs, then current radio interferometers (including EVLA) will struggle to detect them beyond $z=0.6$.

 For HAE we converted the attenuation at H$\alpha$ to that at the UV assuming an SMC extinction curve \citep[$A_{\rm UV}=1.79\,A_{\rm H\alpha}$;][]{gordon03}. 
The comparison with the dust attenuation of GRB hosts and HAEs is less clear. The sample of \citet{villar08} contains much fewer galaxies with very low attenuation compared to the $z\lesssim1$ GRB host sample, but this is not the case for the sample of \citet{sobral09}. In any case, in order for a GRB host sample to be consistent with the HAE sample, the attenuation of the hosts with current limits at $A_{\rm UV}\sim5-7$ mag would need to be very close to these limits.

\subsection{Radio Spectral Indices}
\label{sec:alpha}

The radio spectral index 
of the host of GRB 980425 turned out to be very steep \citep[$\alpha^{8.64}_{4.8}<-1.44$;][]{michalowski09} and was interpreted as a sign of the dominant old stellar population. 
Similarly, for the GRB 021211 host, using our detection at $1.43$ GHz and the upper limit of $45\,\mu$Jy at $8.46$ GHz reported by \citet{fox03} we derive a steep $\alpha^{8.64}_{1.43}<-1.12$.

On the other hand, the spectral index of the GRB 031203 host is flatter $\alpha^{2.37}_{1.39}=-0.53\pm0.50$. 
 This suggests a higher contribution of free--free emission \citep[or synchrotron self-absorption;][]{condon} and, hence, younger stellar population \citep{bressan02,cannon04,hirashitahunt06,clemens08}. 
This value is also consistent within errors  with 
the spectral slopes of $\alpha\sim-0.7$ to $-0.8$ found for star-forming galaxies both local and at high redshifts  \citep{condon,dunne09,ibar10}.

\section{Conclusions}
\label{sec:conclusion}

We present radio continuum data for a sample of 30 GRB hosts including 22 new observations. We detected three targets. The derived limits on the SFRs show that at least $\sim63$\% of the GRB hosts have SFR$\mbox{}<100\,\msunyr$ and that at most $\sim88$\% of GRB hosts  have $A_{\rm UV}<6.7$ mag, i.e.~$A_V<3$ mag. The average flux of non-detected hosts at $z\sim0.5$ sets an upper limit of  
SFR$\mbox{}< 15\,\msunyr$. Using our radio data in conjunction with the rest-frame ultraviolet data we found that the distributions of SFRs and ultraviolet attenuations of GRB hosts are consistent with those of other star-forming galaxies at $z\lesssim1$. This  is  consistent with the hypothesis that GRBs trace cosmic star formation,  but further studies of morphology and metallicities of GRB hosts are required to understand potential biases in this sample.

\acknowledgments 

We wish to thank Joanna Baradziej  for discussion and comments, Jamie Stevens for an extensive help with the ATCA data, Nirupam Roy for conducting GMRT observations, David Sobral for kindly providing the SFRs of H$\alpha$ emitters, Bunyo Hatsukade for kindly providing his ATCA map of GRB 021211,  and Philip Edwards for allocating ATCA Director's Time to our project.

MJM acknowledges the support of the Science and Technology Facilities Council. 
The Dark Cosmology Centre is funded by the Danish National Research Foundation.
AK and DLK are partially supported by NSF award AST-1008353.
LB acknowledges partial financial support from the Spanish Ministerio de Ciencia e Innovaci\'on project AYA2010-21766-C03-01.
JSD acknowledges the support of the Royal Society via a Wolfson Research Merit award, and also the support of the European Research Council via the award of an Advanced Grant.
The Australia Telescope is funded by the Commonwealth of Australia for operation as a National Facility managed by CSIRO.
We thank the staff of the GMRT who have made these observations possible. GMRT is run by the National Centre for Radio Astrophysics of the Tata Institute of Fundamental Research.
 The National Radio Astronomy Observatory is a facility of the National Science Foundation operated under cooperative agreement by Associated Universities, Inc.
The Westerbork Synthesis Radio Telescope is operated by the ASTRON (Netherlands Institute for Radio Astronomy) with support from the Netherlands Foundation for Scientific Research (NWO).
This research has made use of the GHostS database (\urltt{http://www.grbhosts.org}), which is partly funded by Spitzer/NASA grant RSA Agreement No. 1287913;
Jochen Greiner GRB list (\urltt{http://www.mpe.mpg.de/$\sim$jcg/grbgen.html});
the NASA/IPAC Extragalactic Database (NED) which is operated by the Jet Propulsion Laboratory, California Institute of Technology, under contract with the National Aeronautics and Space Administration;
SAOImage DS9, developed by Smithsonian Astrophysical Observatory \citep{ds9};
Karma toolkit \citep[\urltt{http://www.atnf.csiro.au/computing/software/karma/};][]{karma}, 
GRBlog \citep[\urltt{http://www.grblog.org};][]{grblog};
and NASA's Astrophysics Data System Bibliographic Services.

{\it Facilities:} \facility{ATCA}, \facility{GMRT}, \facility{VLA}, \facility{WSRT}.

\end{document}